\documentclass[11pt,twoside]{article}
\usepackage{asp2010}

\resetcounters

\begin{document}

\title{Monitoring Quasar Colour Variability in Stripe 82}
\author{Jesse A. Rogerson,$^1$ Patrick B. Hall,$^1$ Chelsea MacLeod,$^2$ and \v{Z}eljko Ivezi\'{c}$^2$
\affil{$^1$Department of Physics and Astronomy, York University, Toronto, ON M3J 1P3, Canada}
\affil{$^2$Department of Astronomy, University of Washington, Seattle, WA 98195 USA}}

\begin{abstract}
Broad Absorption Line (BAL) trough variability is predominantly due to cloud motion transverse to our line of sight.  The rate at which the variability occurs indicates the velocity of the cloud, which can provide constraints on the cloud's distance from the central source.  This requires detailed spectroscopy during a variability event.  Such spectra have proven elusive, suggesting either the timescale of variability is too short to be caught, or too long to notice until a sufficient amount of time has passed.  Photometric monitoring of BAL quasar colours may potentially be used as an early warning system to trigger time resolved spectroscopic monitoring of BAL variability.  Towards this end, we are analyzing both BAL and non-BAL colour variability using time series photometry from Stripe 82 in the Sloan Digital Sky Survey.
\end{abstract}

\section{Stripe 82 and Statistical Variability}
A small section of sky observed by the Sloan Digital Sky Survey (SDSS) known as Stripe 82 (S82) was specifically targeted for multiple photometric observations.  The stripe covers the area 22h 24m$ < $ RA$ < $ 04h 08m and $|$Dec$| <$ 1.27 degrees, approximating 290 square degrees on the sky.  S82 has been observed over 50 times with revisit time ranging from 3 hours to 8 years.  There are 9275 spectroscopically confirmed quasars in the Stripe 82 SDSS Data Release 7 \nocite{MIK10}({MacLeod} et~al. 2010).  We sorted these into BAL and non-BAL quasars based on the classifications done in \nocite{GJB09}{Gibson} et~al. (2009) and \nocite{AHM11}{Allen} et~al. (2011) (DR5 and DR6, respectively).

Our working hypothesis is that BAL quasars will exhibit higher variability in their colour than non-BAL quasars as a result of the changes in absorption troughs.  We use the time-series nature of the S82 photometry to extract a signal of higher variability among the BAL quasars.  Nearly all BAL quasars exhibit absorption due to {C\,{\sc iv}} 1550, and so we build our preliminary study around this transition.

Using the possible range in blueshift of a BAL trough (-25 000 km s$^{-1}$ to -3000 km s$^{-1}$, as defined in \nocite{WM91}{Weymann} et~al. 1991) and the wavelength range of the SDSS filters, we sort the quasars of S82 into redshift bins based on where the {C\,{\sc iv}} trough will occur.  For example, in the redshift range $1.73 < z < 2.51$, any {C\,{\sc iv}} broad absorption trough will land entirely within the SDSS $g$ filter.  We can therefore use the colour $g-r$ to compare the colour variability of BAL and non-BAL quasars in that redshift range.

\section{Preliminary Results}

To quantify the overall variability of a given colour, we use the $\chi$ distribution $\chi_i = (c_i - c_0)/(\sigma_i)$, where $c_i$ is the $i$th colour in a series, $\sigma_i$ is the observational uncertainty on the colour, and $c_0$ is the mean colour.  In S82, we observe a higher number of large positive and negative $\chi$ values in the BAL colours compared to non-BALs (significance: $10^{-15}$), as well as larger values of the reduced $\chi^2$ statistic.  This indicates that, overall, BAL quasars exhibit larger deviation from their mean colours than non-BAL quasars.

\begin{figure}
\epsscale{1.0}
\plottwo{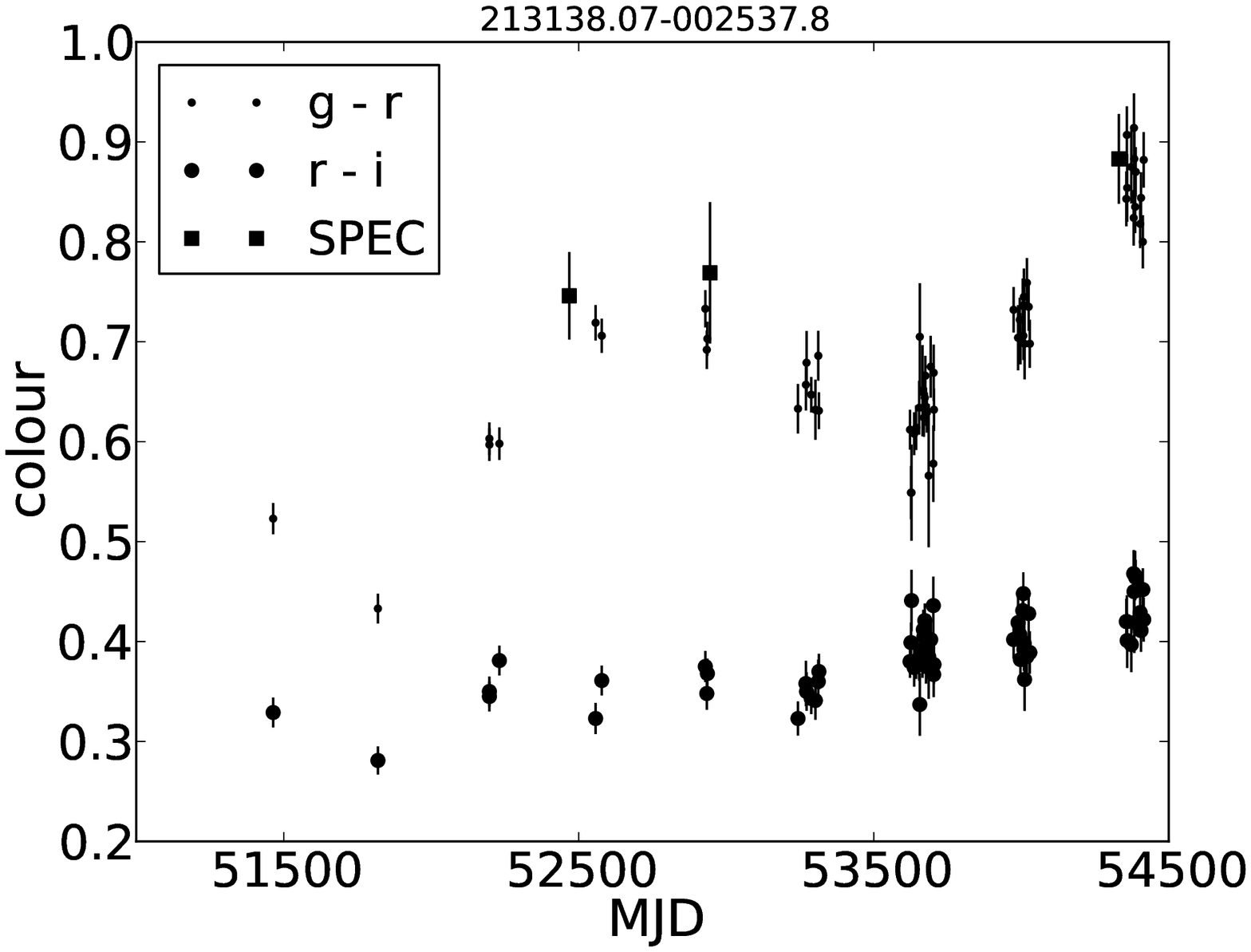}{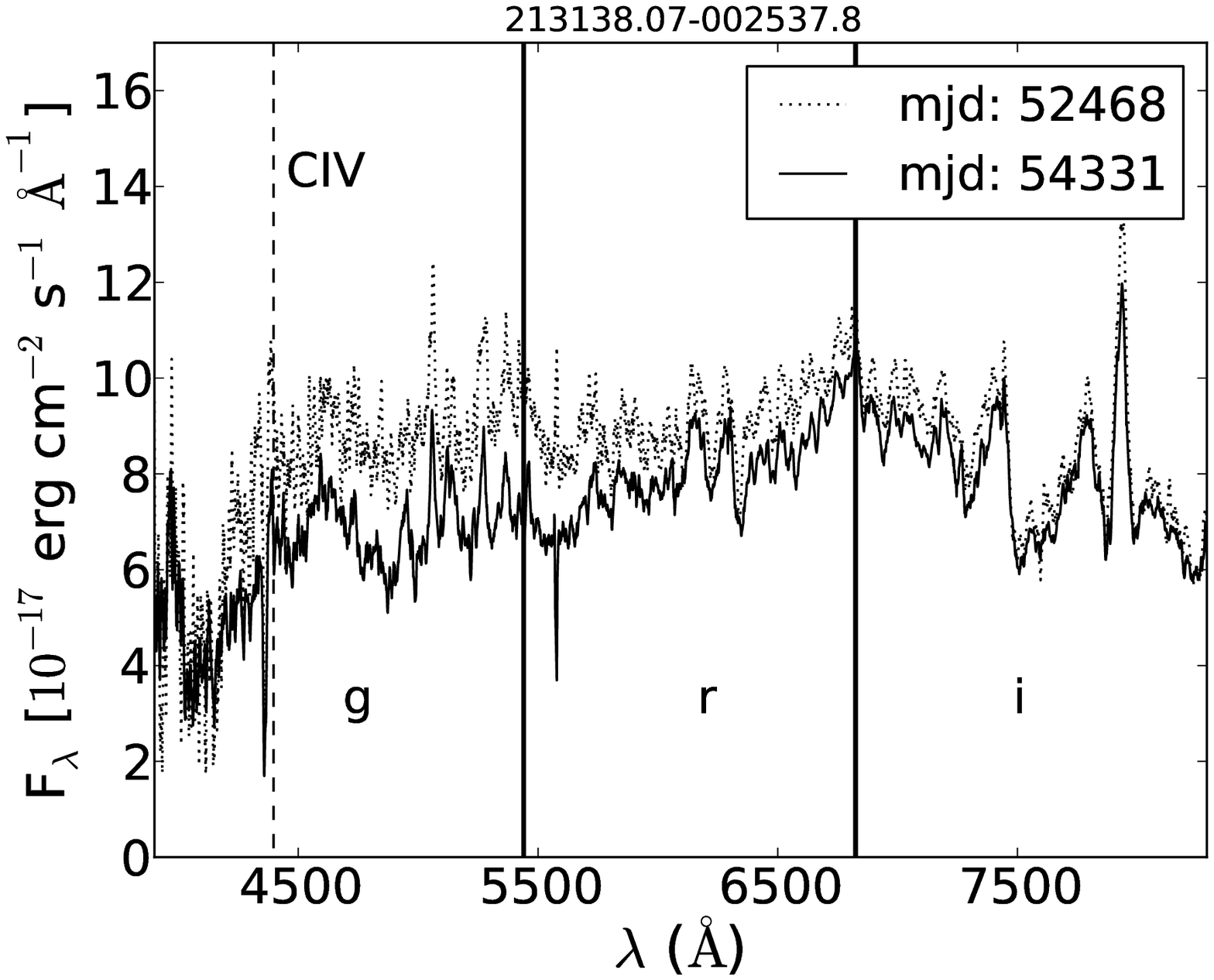}
\caption{SDSSJ 213138.07-002537.8 photometry (left) and spectra (right).}
\label{example}
\end{figure}

As an example of a variable BAL quasar (selected for large reduced $\chi^2$), in Fig. \ref{example}a, we have plotted the time-series colours $g-r$ and $r-i$ (as a comparison), along with estimated colour measurements from spectra taken at the given MJD, of SDSSJ 213138.07-002537.8, a BAL quasar at $z=1.837\pm0.002$.  It is clear that large variations have occured in $g-r$, where the {C\,{\sc iv}} trough is, but not in $r-i$.  In Fig. \ref{example}b, we have plotted the three spectra indicated by the square symbols.  There are clear changes in the {C\,{\sc iv}} trough which contributes to the colour changes observed.  The continuum has also changed by a significant amount.

The S82 photometry provides an excellent testbed to compare colour variability in different types of quasars.  We have found BAL colour to be more variable than non-BAL colour, a feature attributed to the variable nature of absorption troughs.  Such a result hints towards a photometric monitoring program that would trigger spectroscopic followup and coverage when a colour begins to vary.  These data also provide a basis from which to track the colours of non-BAL quasars that may turn into BALs.

\acknowledgements JAR and PBH are supported by NSERC.  We acknowledge support by NSF grant AST-0807500 to the University of Washington, and NSF grant AST-0551161 to LSST for design and development activity.


\begin{thebibliography}{}

\bibitem[{Allen}, {Hewett}, {Maddox}, {Richards}, \&  {Belokurov} 2011]{AHM11}
{Allen}, J.~T., {Hewett}, P.~C., {Maddox}, N., {Richards}, G.~T., \&  {Belokurov}, V. 2011, \mnras, 410, 860.

\bibitem[{Gibson}, {Jiang}, {Brandt}, {Hall}, {Shen},  {Wu}, {Anderson}, {Schneider}, {Vanden Berk}, {Gallagher}, {Fan}, \&  {York} 2009]{GJB09}
{Gibson}, R.~R., {Jiang}, L., {Brandt}, W.~N., {Hall}, P.~B., {Shen}, Y., {Wu},  J., {Anderson}, S.~F., {Schneider}, D.~P., {et al.} 2009, \apj, 692, 758.

\bibitem[{MacLeod}, {Ivezi{\'c}}, {Kochanek},  {Koz{\l}owski}, {Kelly}, {Bullock}, {Kimball}, {Sesar}, {Westman}, {Brooks},  {Gibson}, {Becker}, \& {de Vries} 2010]{MIK10}
{MacLeod}, C.~L., {Ivezi{\'c}}, {\v Z}., {Kochanek}, C.~S., {Koz{\l}owski}, S.,  {Kelly}, B., {Bullock}, E., {Kimball}, A., {Sesar}, B., {et al.} 2010, \apj,  721, 1014.

\bibitem[{Weymann}, {Morris}, {Foltz}, \&  {Hewett} 1991]{WM91}
{Weymann}, R.~J., {Morris}, S.~L., {Foltz}, C.~B., \& {Hewett}, P.~C. 1991,  \apj, 373, 23

\end{thebibliography}

\end{document}